\documentclass[intlimits,twoside,a4paper]{article}

\usepackage{amsmath,amssymb}
\usepackage{graphicx}

\title{Computational physics approaches to light transmission through a twisted nematic liquid crystal slab}
\author{Nadina Gheorghiu and George Y. Panasyuk}

\begin{document}

\maketitle

\begin{abstract}
We consider light propagation through a twisted nematic liquid crystal.
At first, an expression for light transmission is obtained using a rather
intuitive approach. Secondly, an accurate solution for light transmission based on Maxwell's equations 
is derived and compared with the previous one. Both approaches show that when changes in the orientation of the liquid crystal 
optical axis are small on the scale defined by the optical
wavelength (Mauguin limit), the polarization of light approximately follows the optical axis.
At the same time, even in this limit, the simple formula for the light transmittance
mentioned in some monographs on liquid crystal displays is not necessary 
accurate. Conditions under which the two approaches give the same expression for the light transmission are found.
In addition, two numerical methods for finding the light transmittance
are considered. It is demonstrated that the Gauss-Seidel method has much faster convergence rate than the
Euler method. 
\end{abstract}

\section{Introduction}
\begin{figure}[htb]
\vspace{0.5in}
\centerline{\includegraphics{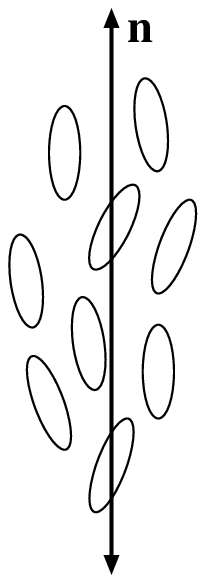}}
\caption{}
\label{Fig1}
\end{figure}

\begin{figure}[htb]
\vspace{0.5in}
\centerline{\includegraphics{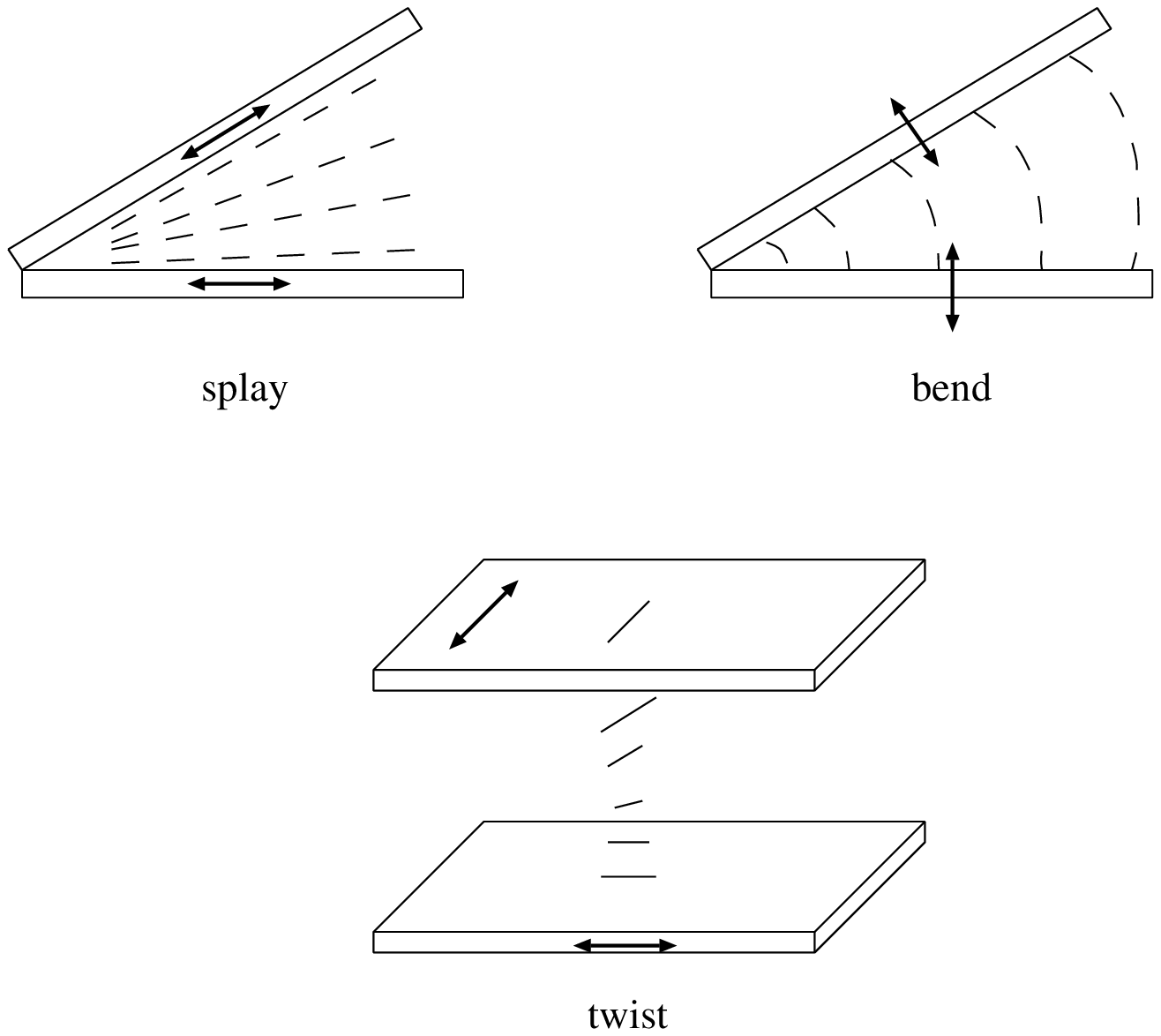}}
\caption{}
\label{Fig2}
\end{figure}

As their name suggests, liquid crystals (LCs), in particular nematic liquid crystals (NLCs) that we consider
here, are mesophases between the liquid and the crystalline states of matter \cite{DeGennes}.
LCs can flow like ordinary fluids, yet they also display an orientational long-range order, which is due to
the anisotropic, rod-like shape of molecules. This \textit{intrinsic anisotropy} of NLCs can be described by
a unit vector \textbf{n}(\textbf{r}), called the
\textit{director}, aligned along the direction of preferred orientation of the molecules at a space point \textbf{r} (Fig. \ref{Fig1}). 
The states \textbf{n} and $-$\textbf{n} are indistinguishable, since the same number of
 molecules orient, on average, along each of these directions.  There are three elastic deformations that
can occur in the bulk of an uniaxial nematic (Fig. \ref{Fig2}). The director alignment can be imposed along a particular
 direction by special treatment of the glass electrodes confining the LC. Assuming no abrupt changes or
 singularities (points or lines), the director deformations can be described within the continuum theory \cite{DeGennes}.
Thus, LCs are not ordinary fluids. Optical patterns of NLCs subjected to electric and magnetic fields show the
 very complex nature and behavior of these systems \cite{Gheorghiu}. The delicate balance between rigidity and fluidity
 makes LCs perfect components of biological systems \cite{Collings,Kleman}. It was found that short DNA oligomers can form LC
 phases by a mechanism responsible for prebiotic formation of DNA molecules on primordial Earth \cite{Nakata}. At higher
 DNA concentration, the higher-order columnar LC phase$^1$ was found to exhibit dendritic growth forms,
 indicative of lower symmetry and a more solid-like ordering. Mostly known for their use in the display industry, 
 LCs are unique in the diversity of their properties \cite{Palffy}. For instance, the existence of LC order in iron-based superconductors was
 experimentally proved by neutron scattering studies that showed the role played by the magnetic degrees of freedom in driving 
 the formation of electronic LCs \cite{Hinkov}. Quantum fluctuations in solid state lattices give rise to electronic LC states, 
 which break the rotational symmetry of the lattice while partially preserving its translational symmetry \cite{Kivelson}.
 
Maxwell's equations are central to the description of light propagation through LCs.
Likewise the case of all other equations from mathematical physics, solutions to Maxwell's equation for two- or three-dimensional problems
have become available only with the advent of fast computers and, especially, computational algorithms. 
Many problems are still not amenable to even the modern computer power. Therefore, the development of both analytical and computational
methods as well as the ability to optimally combine them is essential for the task of finding both accurate and
less time-consuming solutions to physics problems. As Maxwell himself once wrote: ``\ldots the aim of exact science
is to reduce the problems of nature to the determination of quantities by operations with numbers.''

To illustrate these general ideas, we choose here an example from LC optics.
There are two problems that often arise in any LC study.
The first problem is to find the director field \textbf{n}(\textbf{r}) everywhere within the LC cell volume,
which can be done by minimizing the corresponding (Frank-Oseen) free energy \cite{DeGennes}.
The second problem is to calculate light transmittance through the LC cell by solving Maxwell's equations.
Generally, these can be complicated problems and time-effective methods are needed in order to tackle them.
Despite the fact that in the vast majority
of cases only numerical solutions are available, we show here an interesting and, at
the same time, non-trivial example of analytical solution to the problem of calculating the light
transmittance through a LC slab with known (helical) director structure.

\section{Light transmission through a twisted nematic liquid crystal layer}

\begin{figure}[htb]
\vspace{0.5in}
\centerline{\includegraphics{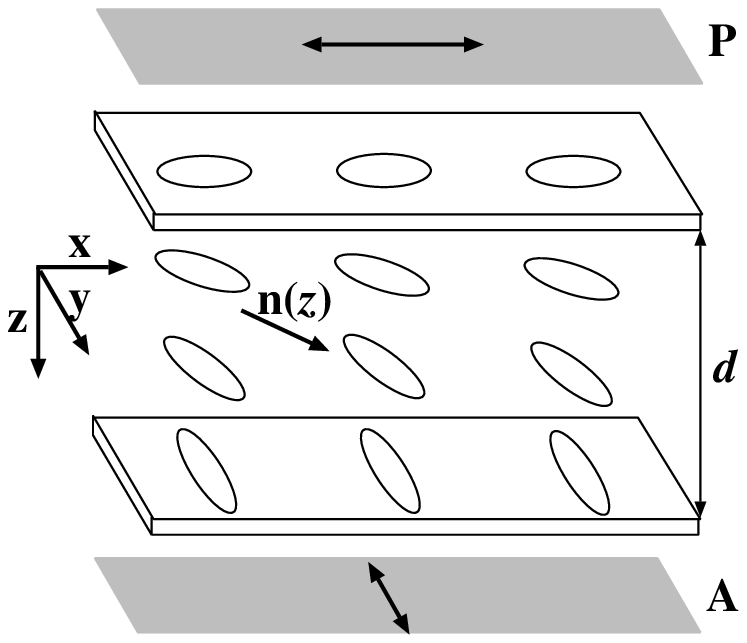}}
\caption{}
\label{Fig3}
\end{figure}

Consider the propagation of an electromagnetic wave of optical frequency $\omega$ that enters along the perpendicular 
direction a LC slab of thickness $d$. 
The bounding plates (Fig. \ref{Fig3}) impose the boundary conditions for the director alignment:
\textbf{n}(0) = $\bf{x}$ and
\textbf{n}($d$) = $\bf{y}$.
Intuitively, one can suggest that in order to minimize its energy
the NLC will acquire a helical structure along the normal direction ${\bf z}$ to the layer:
 \begin{equation}
{\bf{n}}(z) = \cos(qz){\bf{x}} + \sin(qz){\bf{y}},
\label{2D_director}
\end{equation}
where $q = \pi/2d$, $\bf{x}$, $\bf{y}$, and $\bf{z}$ are the unit
vectors along the axes  of the laboratory coordinate system $xyz$, while $z = 0$ is where the electromagnetic wave
enters the LC layer. As one can easily verify, a simple mathematical approach of the problem brings exactly the
same result \cite{DeGennes}.
 Addition of two polarizers (A and P, shown in Fig. \ref{Fig3}) to the LC slab results in what
is widely  known as an essential part of a twisted nematic liquid crystal (TNLC) display.
We assume for simplicity that no electric or magnetic field are applied on the TNCL, apart from the
time-varying electric and magnetic fields in the (electromagnetic) optical wave interacting with the TNLC.
Since the director \textbf{n} cannot follow the high frequency oscillations in the incoming electromagnetic
wave, its helical structure remains stationary.

The next step is to find the amount of light transmitted through the TNLC layer.
After passing through the entrance linear polarizer (P), only the $x$ component of the electric field in the
electromagnetic wave will survive. While advancing through the TNLC layer, the electric field of the wave
will acquire also a $y$ component. The reason for this is that the optical axis (director) ${\bf n}$ rotates
in accordance to Eq. \ref{2D_director}. To describe quantitatively this wave propagation, we assume that
changes in $\bf{n}$ on the scale of the optical wavelength $\lambda$ are small. The imposed condition
$\lambda \ll d$ is satisfied only approximately, as usually $d \geq 5
\mu$m. In this case, we can divide the TNLC layer into $N \gg 1$ thin sheets. Inside each of these sheets,
$\bf{n}$ is a constant vector rotated by a small angle $\delta\phi$ with respect to the preceding sheet. The
electric component
${\bf E}_{m}$ of the electromagnetic field after exiting the $m$-th sheet can be generally written, in  general, as a
vector sum
\begin{equation}
{\bf E}_{m} = E_{m||}{\bf n}_{m} + E_{m\bot}{\bf s}_{m}
\label{E_in_slab}
\end{equation}
where the unit vector ${\bf s}_{m}$ lies in the $xy$-plane, ${\bf s}_{m} \perp {\bf n}_{m}$, and ${\bf n}_{m}$
 is the director in the $m$-th sheet. In this way, ${\bf E}_{m\bot}$ is called the \textit{ordinary} component
 and ${\bf E}_{m\|}$ is called the \textit{extraordinary} component. The two waves propagate with speeds $v_{o}
 = c/n_{o}$ and $v_{e} = c/n_{e}$, respectively, where $n_{o}$ and $n_{e}$ are the ordinary and the
 extraordinary indexes of refraction and $c$ is the speed of light in vacuum. On the other hand, ${\bf E}_{m}$
 can be also written as
\begin{equation}
{\bf E}_{m} = E_{m\|}^{\prime}{\bf n}_{m+1} + E_{m\bot}^{\prime}{\bf s}_{m+1}
\label{E_in_slab_2}
\end{equation}
Equating the two expressions for ${\bf E}_{m}$ (Eqs. \ref{E_in_slab},\ref{E_in_slab_2}) and projecting both
 sides on the ${\bf n}_{m+1}$ and ${\bf s}_{m+1}$ directions, we come up with the following expressions for the
 ordinary and extraordinary components at the exit of the $(m+1)$-th sheet
\begin{equation}
\begin{array}{rcl}
E_{m+1,\bot} \equiv E_{m\bot}^{\prime} & = & E_{m||}\sin(\delta \phi) + E_{m\bot}\cos(\delta \phi)\\
\\
E_{m+1,\|} \equiv E_{m\|}^{\prime}e^{id\Phi} & = & \left[E_{m\|}\cos(\delta \phi) - E_{m\bot}\sin(\delta
 \phi)\right]e^{id\Phi}\\
\label{E_m+1}
\end{array}
\end{equation}
The phase factor $e^{id\Phi}$ appears due to the difference between  $v_{o}$ and $v_{e}$. Here $d\Phi =
 k_{0}(\Delta n)$d$z$, where $k_{0} = 2\pi/d$ and $\Delta n = n_{e}
 - n_{o}$ is called the NLC's \textit{birefringence}. Assuming that d$z = d/N$ is infinitesimally small (when
 $N
 \rightarrow \infty$), one obtains the following system of equations for the ordinary and extraordinary
 components\\
\begin{equation}
\begin{array}{rcl}
\frac{\textstyle dE_{\bot}(z)}{\textstyle {dz}} & \equiv & E_{\bot}^{\prime}(z) = k_{0}E_{\|}(z)\\
\\
\frac{\textstyle dE_{\|}(z)}{\textstyle {dz}} & \equiv & E_{\|}^{\prime}(z) = ik_{0}(\Delta n) E_{\|}(z) -
 k_{0}E_{\bot}(z)
\label{E_deriv}
\end{array}
\end{equation}\\
The light transmission coefficient is determined as the ratio of the transmitted, $I_{T}$, to the incident,
 $I_{I}$, intensity$^7$
\begin{equation}
T = \frac{I_{T}}{I_{I}} = \left(\frac{\left|E_{0T}\right|}{\left|E_{0I}\right|}\right)^{2},
\label{light_tr_def}
\end{equation}
where $E_{0I}$ and $E_{0T}$ are the amplitudes of the electric field at the incidence and at the exit from
the TNLC slab, respectively.
Analytical solution of this system of partial ordinary differential equations (ODEs) with boundary conditions
 $E_{\bot}(0) = 0$ and $E_{\|}(0) = 1$ (to avoid normalization by $|E_{0I}|^2$) can be easily found. Keeping
only  the relevant component, we arrive at:
\begin{equation}
 E_{\bot}(z) = e^{i\pi gz/2d}\sin \left ( {\pi {\sqrt {g^2+1}} \, z\over 2d}\right )
 {\left ( {\sqrt{g^2+1}}\right )}^{-1}
\end{equation}
 Taking into account the presence of the analyzer (A), the light  transmission coefficient will be determined by$^{8}$:
\begin{equation}
T(g) = \left|E_{\bot}(d)\right|^{2} = \frac{\sin^{2}\left[\pi(g^{2} + 1)^{1/2}/2\right]}{g^{2} + 1}
\label{T}
\end{equation}
where $g = 2d(\Delta n)/\lambda$ is a dimensionless parameter. For large values of $g$, which is
consistent with our initial supposition that $d >> \lambda$, light transmittance given by Eq. \ref{T} approaches zero.
 It can be
explained by assuming that the electric field component in the polarized light is aligned approximately along the director. In this
situation, its $x$-component is small, i.e. $|E_x(d)|^2 = |E_{\bot}(d)|^2$ $\simeq 0$. This assumption will be fully proved in 
the next section. As will be shown, the polarization of light propagating through the LC
slab indeed follows the director in the Mauguin  limit \cite{Mauguin} provided that $g >> 1$ (see Eq. \ref{Et_vector}). When
$g = 0$, the result $\left(T(g=0) = 1\right)$ is also correct: if, for example, $d$ = 0 or $\Delta n$ = 0,
 there is no
slab at all or the slab does not change the polarization of the incoming light. However, for intermediate values of
$g$ ($g \leq 1$), Eq. \ref{T} might not be correct, and a more accurate solution presented in the next section
 should
be invoked.

\section{Analytical solution to light propagation based on Maxwell's equations}

We start from the wave equation
\begin{equation}
\nabla (\nabla {\bf{E}}) - {\nabla}^{2}{\bf{E}} = -{1\over c^2}{{\partial}^{2}{\bf{D}}\over {\partial t}^2}
\label{E_wave}
\end{equation}
which can be obtained by eliminating the magnetic component of the electromagnetic field from Maxwell's
 equations \cite{Griffiths,Jackson}.
Here $\bf{D}$ is the displacement vector and, as before, $\bf{E}$ is the electric field component of the wave.
The intrinsic anisotropy of LCs makes the relation between $\bf{D}$ and $\bf{E}$ dependent on the
 orientation of $\bf{E}$ with respect to $\bf{n}$, namely $\bf{D}$ = $\epsilon_{\bot}\bf{E}$ for $\bf{E}$
 $\bot$ $\bf{n}$ and
$\bf{D} = \epsilon_{\|}\bf{E}$ for $\bf{E} \| \bf{n}$. In the general case, $\bf{E}$ is not necessarily
 parallel
to $\bf{D}$ and the relation between $\bf{D}$ and $\bf{E}$ is$^{1,4}$
\begin{equation}
{\bf{D}} = \epsilon_{\bot}{\bf{E}} + \Delta\epsilon\cdot\bf{n}(\bf{n}\bf{E})
\label{DE}
\end{equation}
where $\Delta\epsilon = \epsilon_{\|} - \epsilon_{\bot}$ is the anisotropy in the dielectric constant.
The homogeneity in the $xy$ plane requires $\partial_{x,y}\bf{E} = 0$. Also, because $\bf{n} \bot \bf{z}$,
 $D_{z}
 = 0$. Thus, looking for a solution of Eq. \ref{E_wave} in the form
 ${\bf{E}}(z,t) = {\bf{E}}(z)e^{-i\omega t}$, where ${\bf{E}}(z) = [E_{x}(z),E_{y}(z)]$, Eq.
 \ref{E_wave} reduces to:
\noindent
\begin{equation}
\begin{array}{rcl}
\hskip -0.3 cm -E_{x}^{\prime \prime}(z) & = & k_{0}^{2}\left[\epsilon_{\bot}E_{x} + \Delta\epsilon\cdot
 n_{x}\left(n_{x}E_{x} + n_{y}E_{y}\right)\right] \\
 \\
-E_{y}^{\prime \prime}(z) & = & k_{0}^{2}\left[\epsilon_{\bot}E_{y} + \Delta\epsilon\cdot n_{y}(n_{x}E_{x}
 + n_{y}E_{y})\right]
\label{PDEs_1}
\end{array}
\end{equation}
Taking into account Eq. \ref{2D_director}, Eqs. (\ref{PDEs_1}) can be written as:
\noindent
\begin{equation}
\hskip -0.3cm
\begin{array}{rcl}
 -E_{x}^{\prime\prime}
& = & k_{0}^{2}\bar\epsilon E_{x} + {\textstyle 1\over \textstyle 2}k_{0}^{2}\Delta\epsilon\left[\cos(2qz)E_{x}
 +
 \sin(2qz)E_{y}\right]\\
 \\
-E_{y}^{\prime\prime}
& = & k_{0}^{2}\bar\epsilon E_{x} + {\textstyle 1\over \textstyle 2}k_{0}^{2}\Delta\epsilon\left[\sin(2qz)E_{x}
 -
 \cos(2qz)E_{y}\right]
 \nonumber
\end{array}
\end{equation}
where \, $\bar\epsilon = \left(\textstyle \epsilon_{\|} + \epsilon_{\bot}\right)/2$. Using the Euler
 representation for trigonometric functions,
the previous system of equations can be rewritten in terms of the new variables $E_{1,2}(z) = E_{x}(z) \pm
 iE_{y}(z)$
as \begin{equation}
\begin{array}{rcl}
-E_{1}^{\prime\prime} & = & \frac{\textstyle 1}{\textstyle 2}k_{0}^{2}(\Delta\epsilon) e^{2iqz}E_{2} +
 k_{0}^{2}\bar\epsilon E_{1}\\\\
-E_{2}^{\prime\prime} & = & k_{0}^{2}\bar\epsilon E_{2} + \frac{\textstyle 1}{\textstyle
 2}k_{0}^{2}(\Delta\epsilon) e^{-2iqz}E_{1}
\label{PDEs_3}
\end{array}
\end{equation}
Note that Eqs. \ref{PDEs_3} accurately describe propagation of an electromagnetic wave normally incident to the
 TNLC layer.
Looking for a solution of Eqs. \ref{PDEs_3} in the form
\begin{equation}
E_{1,2}(z) = E_{01,02}e^{i(k \pm q)z},
\label{sol_1}
\end{equation}
we arrive at the following linear equations with respect to $E_{01,02}$:
\begin{eqnarray}
\label{first_eq}
[(k + q)^2 - k_0^2{\bar \epsilon}]E_{01} = {1\over 2}k_0^2\Delta \epsilon E_{02},
\\
\label{second_eq}
{1\over 2}k_0^2\Delta \epsilon E_{01} = [(k - q)^2 - k_0^2{\bar \epsilon}]E_{02},
\end{eqnarray}
and a non-trivial solution for $E_{01,02}$ is only possible if the determinant vanishes, which gives the
 following \textit{dispersion relation}:
\begin{equation}
\label{disp_rel}
(k^2 + q^2 - k_0^2 {\bar \epsilon})^2 = 4q^2k^2 + k_0^4\left( {\Delta \epsilon \over 2}\right )^2
\end{equation}
between $\omega$ (or $k_0 = \omega /c$) and the wave number $k = k(\omega )$ of the propagating wave.

Eq. \ref{disp_rel} has four solutions. Two of these solutions, which we denote by
$k^{(o,+)}=k^{(o,+)}(\omega)$ and $k^{(e,+)}=k^{(e,+)}(\omega)$, correspond to the two waves propagating in the
positive direction of the $z$-axis. The other two, with $k^{(o,-)}=-k^{(o,+)}$
and $k^{(e,-)}=-k^{(e,+)}$, correspond to the two waves traveling in the opposite direction. For convenience,
we introduce the index $m$ standing for ``o'' or ``e'' and the index $s$ corresponding to either ``+'' or ``-''
(forward- or back-propagating wave). For each $k^{(m,s)}$, we can find the ratio
$r^{(m,s)}=E^{(m,s)}_{02}/E^{(m,s)}_{01}$ of the corresponding amplitudes using, for example, Eq.
\ref{first_eq}. Eq. \ref{first_eq} and Eq. \ref{second_eq} become equivalent if $k({\omega })$ satisfies Eq.
\ref{disp_rel}.

A general solution for the electric component $\bf{E}$ of the electromagnetic field in the LC layer will be a
superposition of these four modes and can be presented as:
\begin{eqnarray}
{\bf{E}} = 
\sum_{m,s}\left[{\bf{x}}E_{x}^{(m,s)}(z)+{\bf{y}}E_{y}^{(m,s)}(z)\right] = \sum_{m,s}\left[{\bf{e_{1}}}
E_{01}^{(m,s)}e^{iqz}+{\bf{e_{2}}}E_{02}^{(m,s)}e^{-iqz})\right]e^{ik^{(m,s)}z},
\label{E_sum}
\end{eqnarray}
where ${\bf e}_{1,2}=({\bf{x}} \mp i{\bf{y}})/2$.
Thus, there are four contributions in each sum. As is known, each of
these solutions represent an elliptically polarized wave \cite{Jackson}. The axes of the ellipse, $x'$ and $y'$, are
rotated by an angle $\phi=qz$ with respect the lab coordinates $x$ and $y$. As one can notice immediately from
Fig. \ref{Fig1}, this angle coincides with the director angle. Finally, the ratio $\mu$ of the rotated ellipse semiaxes
(along $x'$ and $y'$) is $\mu^{(m,s)}=\left|(1+r^{(m,s)})/(1-r^{(m,s)})\right|$.

Let us restrict our attention to the visible wavelengths, when $\lambda=2\pi/k\sim0.5$ $\mu$m. This wavelength
region is the most important for the LC physics, in particular for LC displays. Taking into account that
typical values for the helical pitch $p=2\pi/q=4d\sim10\div100$ $\mu$m, the ratio $q/k$ is small. 
It allows for an expansion of
all relations in terms of this ratio and keeping only its first order. With this approximation, Eq.
\ref{disp_rel} gives the following solutions:
\begin{eqnarray}
k^{(o,\pm)} & = & \pm k_{0}n_{o} + O({\delta}^2)\\
\nonumber
k^{(e,\pm)} & = & \pm k_{0}n_{e} + O({\delta}^2),
\label{disp_rel_2}
\end{eqnarray}
where $k_0 = \omega /c$. As follows from Eq. \ref{first_eq}, the corresponding ratios $r^{(m,s)}$ are
determined as:
\begin{eqnarray}
\label{new_param_eq1}
r^{o,\pm}=-1\pm a_{o}\delta+O(\delta ^{2}),\hspace{0.25in} r^{(e,\pm)}=1\pm a_{e}\delta+O(\delta ^{2})
\end{eqnarray}
\begin{eqnarray}
\label{new_param_eq2}
a_{o}=\frac{\textstyle 2n_{o}}{\textstyle n_{o}+n_{e}}, \hspace{0.25in} a_{e}=\frac{\textstyle
2n_{e}}{\textstyle n_{o}+n_{e}}, \hspace{0.25in} \delta=\frac{\textstyle \lambda}{\textstyle 2d\Delta n}=g^{-1}
\end{eqnarray}
As is clear, $\mu^{(o,\pm)}=\left|(1+r^{(o,\pm)}/(1-r^{(o,\pm)})\right|\approx a_{o}\delta/2$.
For small $\delta$, these parameters describe the ordinary waves propagating in the forward or backward
direction, respectively. Both of them are polarized approximately along the $y'$-axis, which is perpendicular
to the director at any point inside the slab. In the second case, $\mu^{(e,\pm)}\approx2/(a_{e}\delta)$
and we have the extraordinary waves approximately polarized along the $x'$-axis, which is parallel to the
director. The electric component ${\bf E}_{t}=\left( E_{tx},E_{ty}\right)$ of the forward-propagating
(transmitted) wave is a superposition of the ordinary and extraordinary waves, traveling in the positive
direction of the $z$-axis. Taking $s = "+"$ in Eq. \ref{E_sum}, the forward-propagating wave can be presented
as:
\begin{eqnarray}
E_{tx}=\frac{\textstyle 1}{\textstyle 2}\sum_{m}\left\{E_{02}^{(m,+)}e^{i\left[k^{(m,+)}-q\right]z}+
E_{01}^{(m,+)}e^{i\left[k^{(m,+)}+q\right]z}\right\}\\
\label{Etx}
E_{ty}=\frac{\textstyle i}{\textstyle 2}\sum_{m}\left\{E_{02}^{(m,+)}e^{i\left[k^{(m,+)}-q\right]z}-
E_{01}^{(m,+)}e^{i\left[k^{(m,+)}+q\right]z}\right\}.
\label{Ety}
\end{eqnarray}
In the same way, the electric component ${\bf E}_{r}=(E_{rx},E_{ry})$ of the back-propagating (reflected) wave
is a superposition of the ordinary and extraordinary waves, traveling in the negative direction of the
$z$-axis. The back-propagating wave can be presented as:
\begin{eqnarray}
E_{rx}=\frac{\textstyle 1}{\textstyle 2}\sum_{m}\left\{E_{02}^{(m,-)}e^{i\left[k^{(m,-)}-q\right]z}+
E_{01}^{(m,-)}e^{i\left[k^{(m,-)}+q\right]z}\right\}\\
\label{Erx}
E_{ry}=\frac{\textstyle i}{\textstyle 2}\sum_{m}\left\{E_{02}^{(m,-)}e^{i\left[k^{(m,-)}-q\right]z}-
E_{01}^{(m,-)}e^{i\left[k^{(m,-)}+q\right]z}\right\}.
\label{Ery}
\end{eqnarray}
Using the definition of the ratio $r^{(m,s)}$, we can express ${\bf{E_{t}}}$ and ${\bf{E_{r}}}$
through only four unknowns $E_{01}^{(m,s)}$:
\begin{eqnarray}
\nonumber
E_{tx}=\frac{\textstyle 1}{\textstyle
2}\sum_{m}E_{01}^{(m,+)}\left\{r^{(m,+)}e^{i\left[k^{(m,+)}-q\right]z}+e^{i\left[k^{(m,+)}+q\right]z}\right\}\\
E_{ty}=\frac{\textstyle i}{\textstyle
2}\sum_{m}E_{01}^{(m,+)}\left\{r^{(m,+)}e^{i\left[k^{(m,+)}-q\right]z}-e^{i\left[k^{(m,+)}+q\right]z}\right\}
\label{Et}
\end{eqnarray}
\begin{eqnarray}
\nonumber
E_{rx}=\frac{\textstyle 1}{\textstyle
2}\sum_{m}E_{01}^{(m,-)}\left\{r^{(m,-)}e^{i\left[k^{(m,-)}-q\right]z}+e^{i\left[k^{(m,-)}+q\right]z}\right\}\\
E_{ty}=\frac{\textstyle i}{\textstyle
2}\sum_{m}E_{01}^{(m,-)}\left\{r^{(m,-)}e^{i\left[k^{(m,-)}-q\right]z}-e^{i\left[k^{(m,-)}+q\right]z}\right\}
\label{Er}
\end{eqnarray}
where $r^{(m,s)}$ are determined by Eqs. \ref{new_param_eq1}, \ref{new_param_eq2} in the considered
approximation.

The corresponding magnetic components, ${\bf{H}}_t$ and ${\bf{H}}_r$, can be easily found from Eqs. \ref{Et},
\ref{Er} and relations
\begin{equation}
\label{H_t,r}
 {\bf{H}}_{t,r} = {i\over k_0} ({\bf x}\partial_zE_{t,ry} - {\bf y}\partial_zE_{t,rx}).
\end{equation}
To properly take into account both the polarizer (P)- and analyzer (A)-LC interfaces, one has to
introduce a plane wave (${\bf{E}}_r^{(P)}$, ${\bf{H}}_r^{(P)}$) reflected from the P-LC interface
and a plane wave (${\bf{E}}_t^{(A)}$, ${\bf{H}}_t^{(A)}$) transmitted through the LC-A interface.
It will bring four additional unknown constants, analogous to $E_{01}^{(n,s)}$, and all eight constants
can be determined from continuity of the tangential ($x$ and $y$) components of the
electric and magnetic fields at both interfaces. As a result, the light transmittance will depend on
the refraction indexes of both polarizer $n_P$ and analyzer $n_A$. At the same time, reflected light was not
considered in the simple derivation of light transmittance discussed in the previous section. It means that
Eq. \ref{T}, strictly speaking, is not correct even in the limit of small $\delta$. Let us,
nevertheless, find conditions under which Eq. \ref{T} is still valid.

In many practical situations, $\Delta n \propto 0.1$ and
reflections are small if one chooses $n_o \leq {\rm Re}(n_{P,A}) \leq n_e$. Indeed, the reflection
coefficient $\bar r$, defined by the relation ${\bf{E}}_r = {\bar r}{\bf{E}}_{inc}$ \cite{Jackson},
will be of the order of ${\bar r} \propto \Delta n/2n_e$ and therefore small. As is also clear, back-propagating wave,
which is a wave reflected from the LC layer itself, will be also of the order of ${\bar r}$, or even of a
higher order. Indeed, the back-scattered wave appears because the optical axis (director) varies with the
$z$-coordinate. In this case we can divide the whole LC slab into a number of thin sheets (as it was
done in the previous section) and consider also reflections at each of these interfaces.
If the birefringence goes to zero, the slab will become homogeneous in the
$z$-direction, these reflections will disappear, and ${\bf{E}}_r$ in Eq. \ref{Er} will vanish.
Thus, to simplify our consideration and make possible the comparison to the results from previous section,
we consider the following situation. Assume that the real part of the either polarizer's and analyzer's index of
refraction is equal to the extraordinary index of refraction, i.e. ${\rm Re}(n_{P,A}) = n_e$. The imaginary part
of $n_{A,P}$ is of the order of $10^{-5} - 10^{-3}$ and therefore negligible. In this case, one
can neglect the reflection of the incident wave from the P-LC interface that is polarized along the director at the
beginning of the LC layer (at $z = 0$). We may also suppose, based on our previous simple consideration, that
the wave propagating through the LC approximately follows the helix. It means that its largest component, which
is polarized along the $y$-axis at the end of the LC layer, will pass the LC-A interface essentially without
reflection. The other (small) component that is polarized along the $x$-axis, will be reflected. Thus, we have
waves that are reflected from both LC layer and LC-A interface that are proportional to the small factor $\bar
r$. These waves will hit the LC-P interface and again will be reflected back to the LC layer. As is clear,
these waves will contribute to the transmitted radiation, but their amplitudes will be proportional to the
small factor ${\bar r}^2$. If we are not interested in these small corrections, we may consider only the
forward-propagating wave (Eq. \ref{Et}), disregarding the back-propagating wave (Eq. \ref{Er}) and also
reflections from the LC-A interface. These simplifying assumptions allow us to compare the result that we will
obtain in this section with the light transmittance derived in the previous section.

Using Eq. \ref{Et}, one can obtain the unknown values $E_{01}^{(n,+)}$ from the following equations:
$E_{tx}(z=0) = 1$ and $E_{ty}(z=0) = 0$. Neglecting $O({\delta}^2)$ contributions, one can easily find:
\begin{equation}
\label{E01_n}
E_{01}^{(o,+)} = {1\over 2}a_e\delta, \,\,\,\, E_{01}^{(e,+)} = 1 - {1\over 2}a_e\delta.
\end{equation}
Substitution of these expressions into Eq. \ref{Et} gives:
\begin{equation}
\label{Et_vector}
{\bf E}_t(z) = {\bf n}(z)e^{ik_0n_ez} + {i\over 2}a_e\delta (e^{ik_0n_oz} - e^{ik_0n_ez})
[{\bf x}\,{\sin}(qz) - {\bf y}\, {\cos}(qz)].
\end{equation}
Corresponding expressions for the magnetic components of the transmitted wave can be obtained from Eq.
\ref{H_t,r}. In particular,
\begin{equation}
\label{Hty}
H_{ty}(z) =  n_e\,{\cos}(qz)\,e^{ik_0n_ez} + {i\over 2}a_e\delta [n_oe^{ik_0n_oz}
- e^{ik_0n_ez}(n_e a_e  - \Delta n)]{\sin}(qz).
\end{equation}
As is known$^{11}$, the light transmittance T through the LC layer can be calculated as
\begin{equation}
\label{T1}
T = {| {\bf S}_d\cdot {\bf z}| \over | {\bf S}_0\cdot {\bf z}|},
\end{equation}
where ${\bf S}_{0,d}$ are the values of the Poynting vector at the beginning ($z = 0$) and at the end ($z =
d$) of the LC slab. In this case, because of the choice of directions for light transmitted through the analyzer and
polarizer, we have:
\begin{equation}
\label{S}
{\bf z} \cdot {\bf S}_{0,d} = {c\over 8\pi} E_{tx}\,H_{ty}^*|_{z = 0,d}.
\end{equation}
Using Eq. \ref{Hty}, one finds that $E_{tx}(0)\,H_{ty}^*(0) = n_e$.
Taking into account that $qd=\pi/2$, \,$e^{\pm \rm i\pi/2}=\pm i$, \,$k_{0}\Delta nd=\pi/\delta$, Eq.
\ref{T1} can be transformed to
\begin{equation}
\label{Tperp}
T \equiv T_{\perp} = {1\over 4n_e}a_e{\delta}^2 (1 - e^{i\pi /\delta })
[a_e n_e - (a_e n_e - \Delta n)e^{-i\pi /\delta }].
\end{equation}
Using Eq. \ref{new_param_eq2} for $a_e$ and observing that
\begin{equation}
\label{dn-appr1}
n_e a_e -\Delta n = {2 n_e n_o\over n_e + n_o} + O({\Delta n})^2,
\end{equation}
one can rewrite Eq. \ref{Tperp} as
\begin{equation}
\label{Tperp1}
T_{\perp} = {a_e^2{\delta}^2n_o\over n_e}{\sin}^2{\pi\over 2\delta}.
\end{equation}
Within our approximation, small corrections of the order of $O({\Delta n})^2 \propto {\bar r}^2$ have been
neglected. In the same way,
\begin{equation}
\label{dn-appr2}
{(n_e + n_o)}^2 = 4n_en_o + O({\Delta n})^2
\end{equation}
and one can represent Eq. \ref{Tperp1} in the form:
\begin{equation}
\label{Tfinal}
T_{\perp} = {\delta}^2{\sin}^2{\pi\over 2\delta}
\end{equation}
which coincides with the result (Eq. \ref{T}) found in the previous section for the case of large $g =
{\delta}^{-1}$ (Mauguin limit). As is also clear from Eq. \ref{Et_vector}, in the limit of small $\delta=1/g$
the polarization of light indeed approximately follows the director (optical axis) $\bf{n}$.

\section{Computational}

As we have already discussed, Eq. \ref{T} for the light transmittance is not accurate. Using the
analytical approach based on Maxwell's equations, one can obtain an
expression for the light transmittance valid for all $g$. However, for vast majority of situations,
when the director does not obey
simple formulas like (\ref{2D_director}) or for the case of oblique incidence of an electromagnetic wave,
it is not possible to derive an accurate analytical expression for the light transmittance. For example, in
the considered TN mode and for a nonzero electric field, the light
transmittance can be obtained only by numerical solving of the corresponding equations. This is even
more true for cases when the director $\bf{n}=\bf{n}(\bf{r})$ has a three-dimensional structure.
varies along all space directions. Despite the lack of analytical formulas,
there is usually an excellent correspondence between
numerical solutions to these problems and
measurements of the light transmittance provided that the numerical
solver is accurate and fast enough.
This often enables one, for example, to optimize the performance of
a LC display avoiding thousands or sometimes even millions of costly measurements.

In this Section we show two numerical approaches, Euler and Gauss-Seidel, for solving ODEs like
Eqs. \ref{E_deriv} and demonstrate the advantage of the latter one.
Consider a system of $N$ first-order
linear ODEs to determine an $N$-component vector $\textbf{E}(z)$ on an interval $a \leq z \leq b$ with known
$\textbf{E}\textnormal(a)$. This system can always be written in the matrix form
\begin{equation}
\textbf{E}^{\prime}(z) = \frac{\textstyle d}{\textstyle dz}\textbf{E} = A\cdot\textbf{E},
\label{matrix_form}
\end{equation} 
where $A \equiv A(z)$ is a known matrix. For our system (\ref{E_deriv}), all matrix coefficients are $z$
independent: $A_{11} = 0$, $A_{12} = k_{0}$, $A_{21} = -k_{0}$, and $A_{22} = ik_{0}\Delta h$.
The Euler method for numerically solving Eq. \ref{matrix_form} consists of a step-by-step application of
the formula
\begin{equation}
\textbf{E}(z + dz) = \textbf{E}(z) + A(z)\cdot\textbf{E}(z)dz = \left[I + A(z)dz\right]\textbf{E}(z),
\label{Euler}
\end{equation}
where $I$ is the $N \times N$ unit matrix and $dz = d/N$ is the mesh step.
Alternatively, the Gauss-Seidel (GS) method consists of solving the equation
\begin{equation}
\textbf{E}(z + dz) = \textbf{E}(z) + \frac{\textstyle 1}{\textstyle 2}\left[A(z + dz)\cdot\textbf{E}(z + dz) + A\cdot\textbf{E}(z)\right]dz
\label{Gauss-Seidel}
\end{equation}
with respect to $\textbf{E}(z + dz)$ for each step.
Clearly, the coefficient at $dz$ in Eq. \ref{Gauss-Seidel} is the average derivative between mesh points $z$
and $z + dz$ instead of just $\textbf{E}^{\prime}(z)$ as in the Euler method. Let us show that this substitution
dramatically improves the accuracy of solving Eq. \ref{matrix_form}. Indeed, Eq. \ref{Gauss-Seidel} gives
\begin{equation}
\textbf{E}(z + dz) = \left[I - \frac{\textstyle 1}{\textstyle 2}A(z + dz)dz
\right]^{-1}\left[I + \frac{\textstyle 1}{\textstyle 2}A(z)dz
\right]\textbf{E}(z)
\label{improve}
\end{equation}
Assuming that $dz$ is small enough, expansion of the right-hand side of Eq. \ref{Gauss-Seidel}
in powers of $dz$ and substitution of $A(z + dz)$ by $A + A^{\prime}(z)dz + O(dz^2)$ gives for the GS method:
\begin{equation}
\textbf{E}(z + dz) = \left\{I + A(z)dz +
\frac{\textstyle 1}{\textstyle 2}\left[A^{\prime}(z) + 
A^2(z)\right]dz^2 + O(dz^3)\right\}\textbf{E}(z)
\label{GS_new}
\end{equation}
On the other hand, expansion of ${\bf{E}}(z + dz)$ in Taylor series gives:
\begin{equation}
\textbf{E}(z + dz) = \textbf{E}(z) + \textbf{E}^{\prime}(z)dz + \frac{\textstyle 1}{\textstyle 2}\textbf{E}^{\prime\prime}(z)dz^2 + O(dz^3)
\label{Taylor}
\end{equation}
Using Eq. \ref{matrix_form}, one can find
that $\textbf{E}^{\prime\prime}(z) = A^{\prime}\cdot\textbf{E}^{\prime} + A\cdot\textbf{E}^{\prime} = A^{\prime}\textbf{E} + A^2\textbf{E}$,
and Eq. \ref{Taylor} can be rewritten as:
\begin{equation}
\textbf{E}(z + dz) = \left\{I + A(z)dz + \frac{\textstyle 1}{\textstyle 2}\left[A^{\prime}(z) + A^2(z)\right]dz^2 + O(dz^3)\right\}\textbf{E}(z)
\label{GS_last}
\end{equation}
Comparing Eq. \ref{Euler} and Eq. \ref{GS_new} with Eq. \ref{GS_last},
one finds that the error of computing $\textbf{E}(z + dz)$ by the GS method that properly accounts for $O(dz^2)$
corrections is $O(dz^3)$. The same error for the Euler method is $O(dz^2)$. Thus, the GS method is one order
of magnitude more accurate than the Euler method.

\begin{figure}[htb]
\vspace{0.5in}
\centerline{\includegraphics{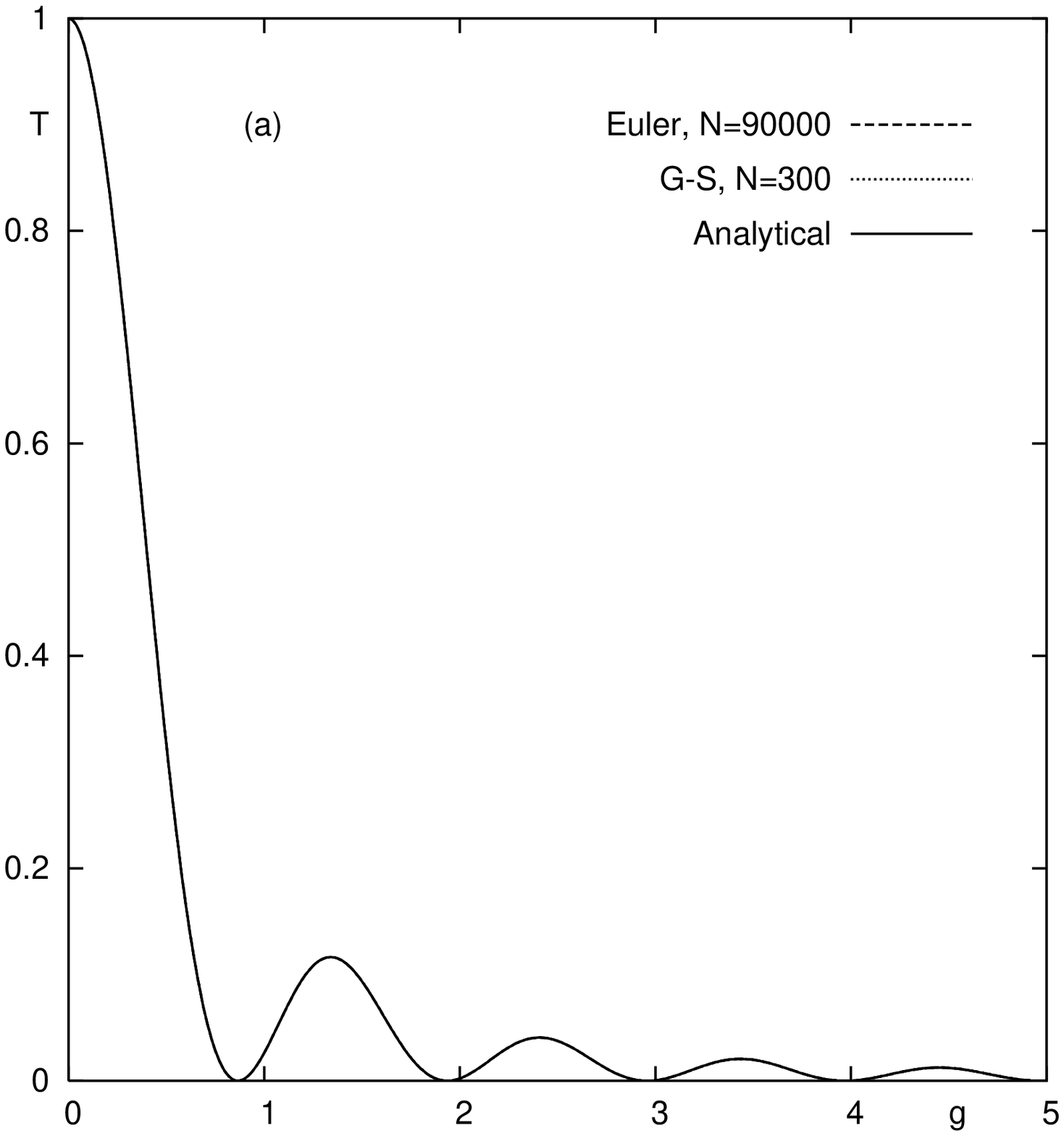}}
\caption{}
\label{Fig4}
\end{figure}

\begin{figure}[htb]
\vspace{0.5in}
\centerline{\includegraphics{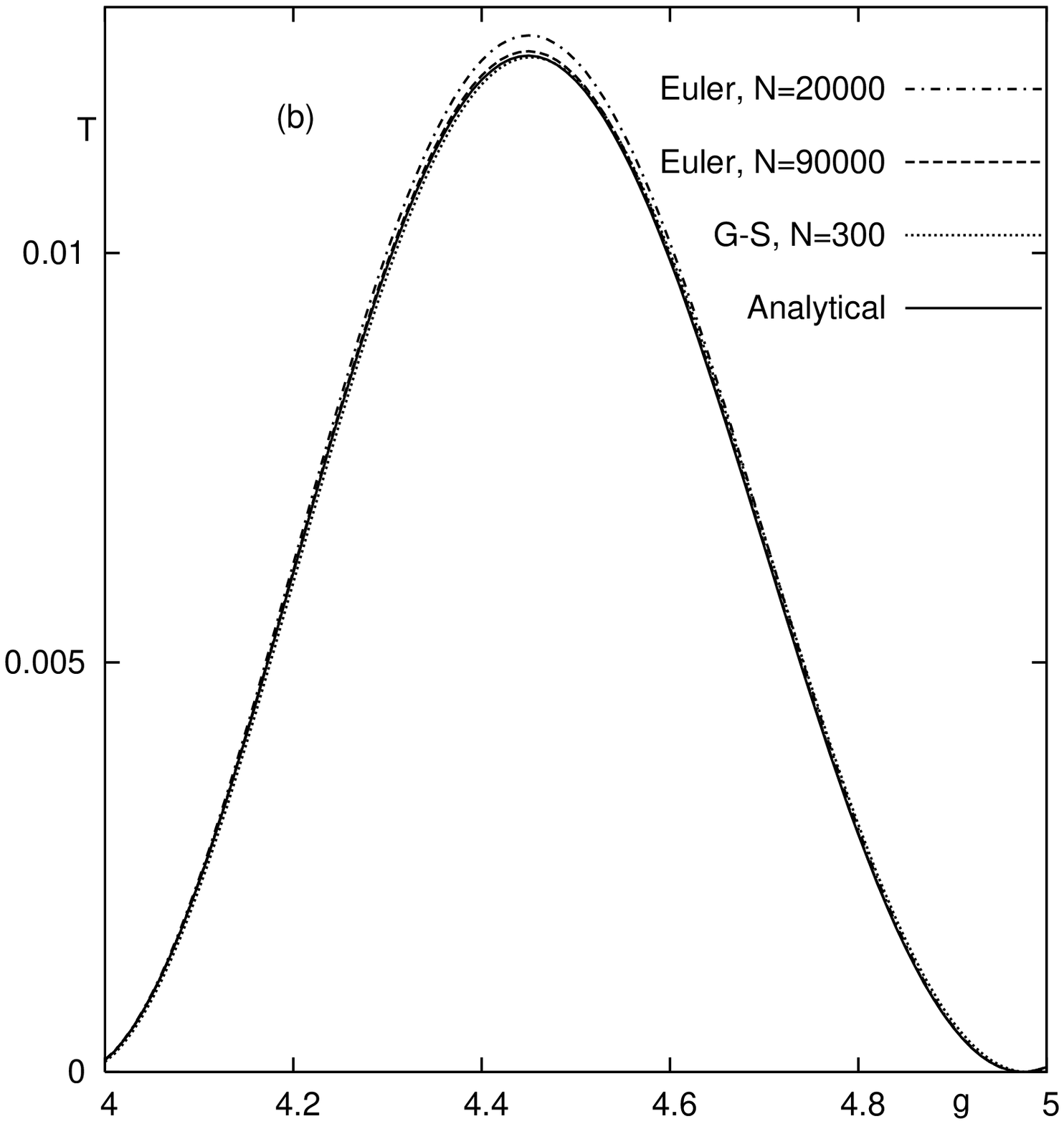}}
\caption{}
\label{Fig5}
\end{figure}

Fig. \ref{Fig4,Fig5} shows results of numerical integration of Eqs. \ref{E_deriv}  by the two methods and their
comparison with analytical solution (8). Fig. \ref{Fig4} contains numerical solutions
obtained by the GS method for $N = 300$ iterations and by the Euler method for $N = 90000$ iterations, when they practically coincide with
the analytical curve. Fig. \ref{Fi5} clearly demonstrates the much better accuracy and faster convergence to
the analytical solution when using the GS method as compared to using the Euler method. In a more accurate
sense, we define characteristic deviations of the numerically computed curves from the analytical one as
$\Delta_{N}^{GS} \equiv {\rm max}\left|I_{N}^{GS} - I^{analytical}\right|$ and $\Delta_{N}^{Euler} \equiv {\rm
max}\left|I_{N}^{Euler} - I^{analytical}\right|$ on the interval $4 \leq g \leq 5$, where these deviations are
the largest. We found that $\Delta_{N=300}^{GS} \simeq 2\times10^{-5} < \Delta_{N=90,000}^{Euler} \simeq
5\times10^{-5}$, and $\Delta_{N < 90,000}^{Euler}$ are increasingly larger. The GS method can be successfully used
to solve significantly more complicated three-dimensional problems in LCs \cite{Panasyuk} or elsewhere.

\section{Discussion and Conclusions}

We have considered light propagation and its transmission through an anisotropic medium: a TNLC.
First, an intuitive analytical solution for the light transmission based on dividing the LC slab into a
large number of thin sheets was derived. Subsequently, this solution was checked against another approach based
on accurate solution for Maxwell's equations applied to light propagation through the LC slab. As shown, the simple formula (Eq. \ref{T}) is
not generally correct even in the Mauguin limit when the parameter $g = 2d\Delta n /\lambda$ is large. Indeed, as shown in Section III, there is a back-scattered wave originated by the change of the optical axis (LC director) in the direction of light propagation and by the LC birefringence
$\Delta n = n_e - n_o \neq 0$. In this case we have two waves polarized along and perpendicular to the
director, respectively. These waves have different indexes of refraction, $n_e$ and $n_o$, which makes it impossible to
match them simultaneously with the real part of the refractive index of the polarizer or  analyzer. We found that the error of neglecting this contribution is of the order of $\left(\Delta
n/n_e\right)^2$ and is small when $\Delta n \ll 1$. In this situation, one can neglect it, and the resulting
simple formula for the light transmittance  is restored. 
In the final part of the paper, two numerical approaches for solving differential equations were
considered. The advantage of the computational Gauss-Seidel method over the Euler approach was shown. 

Our approach is trying to clarify particular formulas introduced by books like \cite{Khoo} that are mostly oriented on
technical applications and frequently cite formulas without any derivations. At the same time, monographs like$^1$ are
usually dealing only with fundamental aspects of light propagation and do not
contain neither formulas for light transmittance through the LC slab nor
any details on their technical applications. Thus, our approach to a particular case of the LC physics 
bridges the gap between the two.\\

\newpage
\centerline{\Large List of figure captions}
\vspace{0.25in}
\noindent
FIG. 1. The arrangement of molecules in an uniaxial nematic liquid crystal is described by a unit vector \textbf{n} known 
as a the \textit{director}.

\noindent
FIG. 2. Three types of deformation occurring in uniaxial nematics: splay, bend, and twist.

\noindent
FIG. 3. Twisted nematic liquid crystal layer of thickness $d$ between polarizer P and analyzer A. Inside the
layer, \textbf{n} has a helical structure.

\noindent
FIG. 4. Comparison of numerical results obtained by the Gauss-Seidel method for $N = 300$ and by the Euler method
for $N = 90000$, respectively, to the analytical curve. 

\noindent
FIG. 5. Numerical results show clearly the much better accuracy and
faster convergence to the analytical solution when using the Gauss-Seidel method as compared to using the Euler
method.

\newpage
\begin{thebibliography}{99}
\bibitem{DeGennes}P.G. De Gennes and J. Prost, \textit{The Physics of Liquid Crystals}, 2nd ed.
(Oxford University Press, New York, 1993).
\bibitem{Gheorghiu} N. Gheorghiu and J.T. Gleeson, ``Length and speed selection in dendritic growth of electrohydrodynamic
convection in a nematic liquid crystal'', Phys. Rev. E \textbf{66}, 051710 (2002).
\bibitem{Collings}P.J. Collings, \textit{Nature's delicate phase of matter}, 2nd ed. (Princeton University Press, Princeton, 2002).
\bibitem{Kleman}M. Kleman and O.D. Lavrentovich, \textit{Soft Matter Physics: An Introduction} (Springer, New York, 2002).
\bibitem{Nakata}M. Nakata, G. Zanchetta, B.D. Chapman,, C.D. Jones, J.O. Cross, R. Pindak, T. Bellini, N.A. Clark, 
``End-to-end Stacking and Liquid Crystal Condensation of 6-to-20-Base Pair DNA Duplexes'', Science \textbf{318}, 1276 (2007).
\bibitem{Palffy}P. Palffy-Muhoray, ``The diverse world of liquid crystals'', Physics today \textbf{60}(9), 54 (2007).
\bibitem{Hinkov}V. Hinkov, D. Haug, B. Fauqu\'{e}, P. Bourges, Y. Sidis, A. Ivanov, C. Bernhard, C.T. Lin, B. Keimer,
\textit{Electronic Liquid Crystal State in the High-Temperature Superconductor YBa$_{2}$Cu$_{3}$O$_{6.45}$}, Science \textbf{319}, 597 - 600 (2007).
\bibitem{Kivelson}S.A. Kivelson, E. Fradkin \& V.J. Emery, \textit{Electronic liquid-crystal phases of a doped Mott insulator},
Nature \textbf{393}, 550 - 553 (1998).
\bibitem{Griffiths}D.J. Griffiths, \textit{Introduction to Electrodynamics}, 
3rd ed. (Prentice Hall, New Jersey, 1999).
\bibitem{Khoo}I.C. Khoo and S.T. Wu, \textit{Optics and nonlinear optics of liquid crystals} (World Scientific, Singapore, 1993).
\bibitem{Mauguin}C. Mauguin, Bull. Soc. Fr. Min\'{e}r. Crystallogr. \textbf{34}, 3 (1911).
\bibitem{Jackson}J.D. Jackson, \textit{Classical Electrodynamics}, 3rd ed., p. 300 (Wiley, New York, 1999).
\bibitem{Bohren}C.F. Bohren, D.R.Huffman, \textit{Absorption and scattering of light by small particles} (Wiley, New
York, 1998).
\bibitem{Panasyuk}G. Panasyuk, J. Kelly, E.C. Gartland, and D.W. Allender, ``Geometrical optics approach
in liquid crystal films with three-dimensional director variations'', Phys. Rev. E \textbf{67}, 041702 (2003).
\end{thebibliography}
\end{document}